# Generation of tunable, high repetition rate optical frequency combs using on-chip silicon modulators

KP Nagarjun,[1] Vadivukarassi Jeyaselvan,[1] Shankar kumar selvaraja,[1,2] and V R Supradeepa[1,*]

[1]*Centre for Nano Science and Engineering, Indian Institute of Science, Bangalore 560012, India*
[2]*shankarks@iisc.ac.in*
*\*supradeepa@iisc.ac.in*

**Abstract:** We experimentally demonstrate tunable, highly-stable frequency combs with high repetition-rates using a single, charge injection based silicon PN modulator. In this work, we demonstrate combs in the C-band with over 8 lines in a 20-dB bandwidth. We demonstrate continuous tuning of the center frequency in the C-band and tuning of the repetition-rate from 7.5GHz to 12.5GHz. We also demonstrate through simulations the potential for bandwidth scaling using an optimized silicon PIN modulator. We find that, the time varying free carrier absorption due to carrier injection, an undesirable effect in data modulators, assists here in enhancing flatness in the generated combs.

## 1. Introduction

Frequency combs have been extensively used in applications like WDM optical communications [1], astronomy [2,3], spectroscopy and metrology [3, 4] and sensing [5]. Conventionally, frequency combs are generated using mode-locked lasers or bulk optical modulators. Combs generated in this manner utilize bulky setups, have fixed, low repetition rates and require expensive feedback electronics [6], making them unsuitable for several applications including optical communications. Recently, several on-chip frequency comb generation techniques have been proposed to overcome these drawbacks [7]. These on-chip schemes have the key advantage of being compact, portable and can be mass-produced.

On-chip comb generation techniques in the C-band have typically been investigated in Silicon Nitride resonators, where cascaded four-wave mixing is used to generate Kerr combs [7–11] . These combs, though having a large number of lines [11] depending on the quality factor of the resonator, typically have fixed repetition rates. Further, the input laser has to be locked to a mode of the resonator and any drift of the laser or the resonator over time has to be compensated. In the absence of such stabilization, it is known that the comb ceases to exist. It is strongly desired to have an integrated, high repetition rate, optical frequency comb which is very stable over an extended period of time with the ability to easily change the center frequency as well as the repetition rate. In this work, we demonstrate such a source.

The comb generation mechanism is based on Strong phase modulation. Phase modulating an external continuous wave (CW) laser with an RF sinusoid can be used to generate multiple comb lines which results from the generation of harmonic side-bands [12]. Comb lines generated this way can have tunable repetition rates based on the frequency of the driving RF sinusoid and a tunable center frequency based on the frequency of the CW laser. This technique of comb generation along with envelope shaping has been extensively demonstrated with bulk Lithium-Niobate modulators. However, with phase modulation alone, frequency combs generated this way result in combs with missing lines and poor flatness and require cascading of a series of phase and intensity modulators with complex driving schemes to achieve flat combs [13,14]. Recently, this idea of using strong phase modulation for comb generation was investigated using Silicon-Organic Hybrid (SOH) modulators in an intensity modulator configuration [15]. Since the electro-optic effect in Silicon is weak, an organic cladding was utilized to obtain the required phase modulation using SOH modulator configuration. A flexible and tunable property of the optical material could potentially increase the frequency of operation beyond some of the non-linear crystal. However, the following two disadvantages could weight against SOH. Firstly, the polymer should be polled at very high-voltage at an elevated temperature followed by controlled cooling, which is not feasible in a highly integrated circuit. Secondly, the back-end needs to be opened to access the Silicon waveguide, which could lead to long-term ion-diffusion and subsequent degradation of the Si device layer. In this work, we utilize a charge-injection based Silicon phase modulator fabricated using a standard

Silicon photonics foundry process for the comb generation. Unlike, the SOH platform the devices require no post-processing and can accommodate multiple devices with varying specifications. Since the fabrication is done using mature technology, it allows seamless integration with driver and control electronics, unlike other platforms.

Silicon based charge injection modulators, have become a popular means by which high speed data (several 10's of Gbps) is modulated onto an optical carrier [16]. These devices work by changing the free carrier dispersion in a Silicon optical waveguide through charge injection from doped regions and the change in refractive and absorptive indices is governed by the Soref-Bennett relation [17]. Apart from having a small device footprint it is compatible with standard CMOS fabrication technology and is thus easily integrated with other Silicon photonic devices.

In typical micro-ring based comb generators based on four wave mixing which is the more common technique for generation of integrated combs [7–11], combs with bandwidths exceeding 650 nm have been demonstrated [11] . However, only coarse tuning of the center frequency is possible due to the periodic resonances of the ring. Further, they have a fixed and very high repetition rate (for example 189.2 GHz in [11]). This is dictated by requirement of fixed and short lengths of the micro-ring resonators necessary to obtain the high Q factors required for nonlinear generation. This makes such comb generation systems unsuitable for applications which require agility in the form of tunable center frequency and tunable, reasonable repetition rates. An example would be comb generation for super-channel based optical communications. Here, we demonstrate combs whose center frequency can be both coarse (~nm) and fine (~pm) tuned, which makes the combs generated in this manner very attractive on-chip DWDM and super-channel sources. The combs generated here also have the additional advantage of a tunable repetition rate. In this work, we demonstrate repetition rate tuning between 7.5 to 12.5 GHz. However, it is to be noted that the tunability of the repetition rate is only limited by the bandwidth of the modulator and can be further reduced as desired. In each case, we demonstrate ~8 lines in a 20dB bandwidth. Further, owing to the fact that this technique does not employ resonant structures, this system is very robust to thermal drift. In our experiments, we observed highly stable combs that remained unaltered even during continuous operation of over 12 hours.

In this work, we also investigate bandwidth scaling through electro-optic simulations of an optimal Silicon PIN modulator. A PIN modulator is chosen for the simulations since lower linear loss in comparison to PN modulators enable longer modulators which provide greater phase modulation and thus more lines. We find that free carrier absorption due to carrier injection shapes the amplitude to enable high quality combs from a single modulator instead of a cascade of intensity and phase modulators as needed in Lithium-Niobate equivalent [13]. We demonstrate that over 37 lines is possible in a 1cm device with a flatness of the spectrum better than 5dB. We further compare our simulation to a pure phase modulation and observe that our Si PIN-based design produces comb lines with a flattened spectrum without missing lines in contrast to the pure phase modulation case.

## 2.  Experimental results

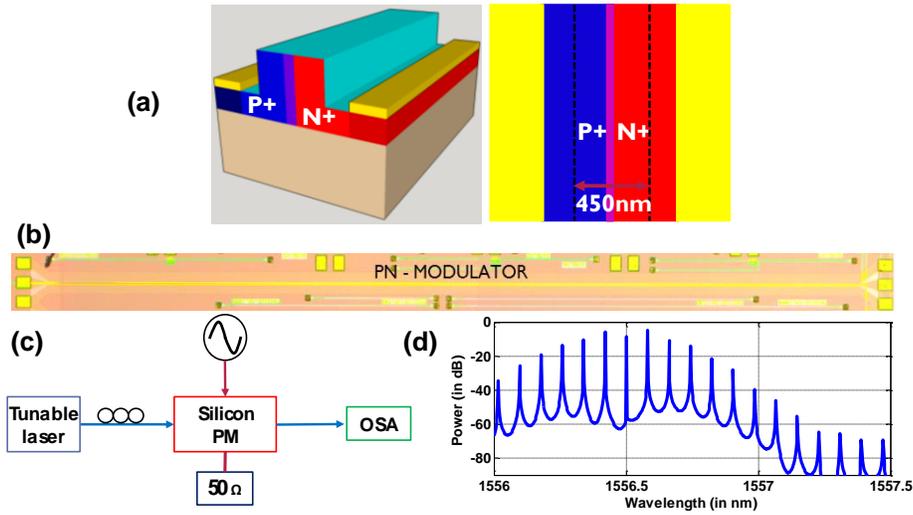

Fig. 1. (a) Side view and top view of the designed device (b) Microscope image of device used for comb generation. (c) Setup for comb generation. (d) Simulated combs from the setup at 10GHz repetition rate.

The devices were fabrication through Europractice multi-project wafer run [18]. We designed and customized a 4.5 mm long travelling-wave PN-phase modulator in a standard 220 nm -SOI wafer. The Silicon rib waveguide width is 450 nm with an etch-depth of 160nm and shoulder height of 60 nm (Fig. 1(a), Fig. 1(b)). An average symmetric doping concentration of $10^{18}/cm^3$ in the N and P regions was chosen for the entire length of the device to prevent excessive absorption due to high doping and simultaneously provide sufficient refractive index modulation. The modulator employs a co-planar travelling wave electrode design. An RF-drive from a tunable microwave source (upto 20GHz) and ~7.8Vpp is used to drive one end of the phase modulators while the other end is terminated with a high frequency 50-ohm load. This method of comb generation does not require a DC bias since it's based on a pure phase modulation based process.

Figure 1(c) shows the schematic of the comb generation setup. A tunable C-band laser with100 kHz linewidth is used as the input. On-chip grating couplers are used in conjunction with single mode fibers to couple light into/out of the modulator. The fiber to waveguide loss in this setup is found to be 2.5dB. The polarization of the input light is optimized using a polarization controller such that maximum power is coupled through the grating coupler. Figure 1(d) shows the simulated comb at 10 GHz repetition rate.

Figures 2(a)-(d) shows the experimental results with increasing RF power inputs for a center frequency of ~1557.84 nm. As expected, we observe an increase in the number of comb lines with the RF power. For an RF power of 27dBm corresponding to ~7.8Vpp, we observe 8 comb lines in a 20dB bandwidth.

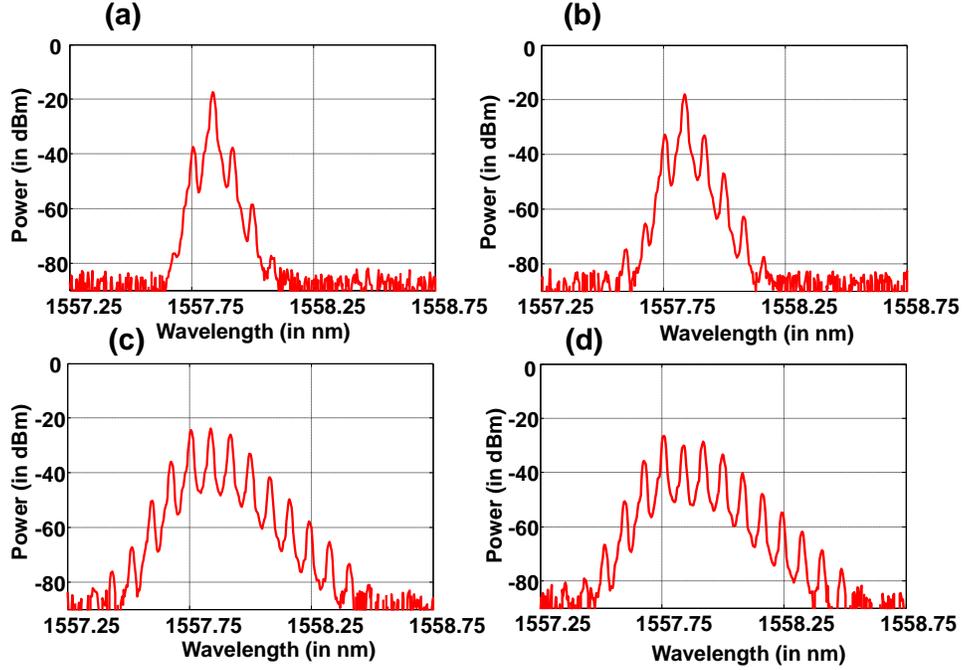

Fig. 2. Frequency Comb generation in a travelling-wave Silicon PN-modulator for various RF-powers at 10 GHz (a) +14dBm (b) +20 dBm (c) +23 dBm (d) +27 dBm.

The center frequency of the comb can be continuously tuned, with the tunability only being limited by the bandwidth of the input-output grating couplers which can routinely have 1dB bandwidths spanning the C-band. Here, we show a representative plot of center frequency shifting by ~1.2 nm (Fig. 3).

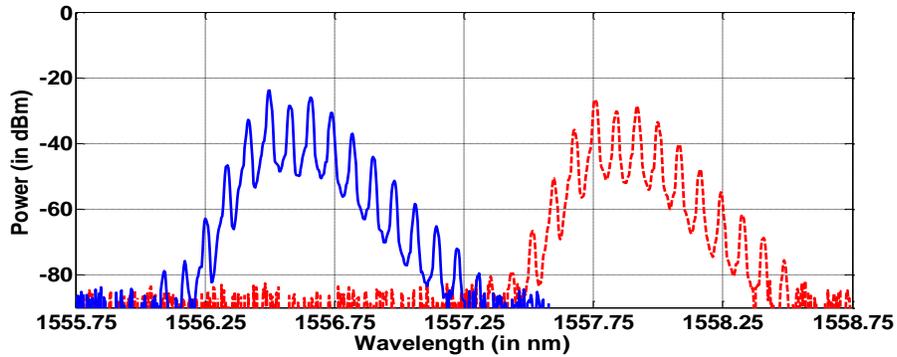

Fig. 3. Tuning of the center frequency of the comb from ~1556.6 nm (blue) to ~1557.84nm (red)

We also performed experiments of tuning of repetition rate of the comb from 5GHz to 15GHzand representative plots of the comb at 7.5GHz and 12.5GHz are shown (Fig. 4). Beyond 12.5 GHz, we observe a limitation that the number of comb lines decrease with increasing frequency. The number of lines reduces to 4 in a 20dB bandwidth at 15GHz.

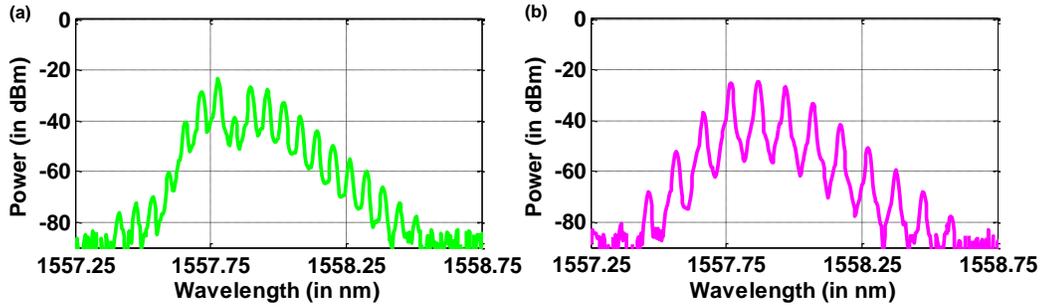

Fig. 4. Comb Repetition rate tuning(a) 7.5GHz (b) 12.5GHz.

We also observed that these combs are stable over very long durations. In our experiments, the combs were allowed to run freely for over 12 hours and were unaltered in their shape.

The measured linear absorption loss for the modulator is ~16dB. It is observed that the linear absorption loss does not depend on RF drive frequency. We also observe that the maximum CW power that can be applied on the device is 75mW beyond which the net output power saturates due to nonlinear absorption. There is also significant contribution to this from the passive input/output coupling waveguides which together is ~3.3 mm.

The method for comb generation used in this work, namely strong phase modulation in a single pass modulator is anticipated to generate a coherent comb as has been observed in Lithium-Niobate modulator based systems. In order to evaluate any degradation of coherence or linewidth across the generated comb, we performed the following experiment with the comb at 10GHz repetition rate. Starting from the center line, we choose neighboring pairs of comb-lines using a programmable spectral filter and generated their respective beat notes using a high-speed photodetector and RF spectrum analyzer. Figure 5 shows the beat signal spectra as we move across the comb to one side (right).

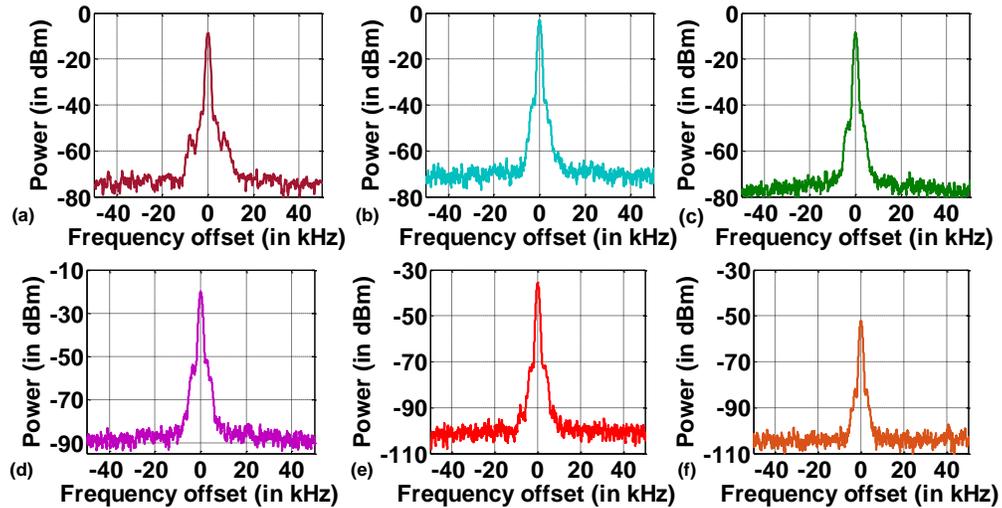

Fig. 5. Beat notes between successive comb lines. Frequency offset of 0 set to the repetition rate of the comb. (a) center and 1st line to the right (b) 1st and 2nd line (c) 2nd and 3rd line (d) 3rd and 4th line (e) 4th and 5th line (f) 5th and 6th line to the right respectively.

Here, the frequency offset of '0' is fixed to be the comb repetition rate. We consistently measure very similar beat notes across the spectrum with a sub-kHz bandwidth. This is substantially smaller than the natural laser linewidth of 100 kHz. At these bandwidth levels, we see that there is no degradation of the coherence of the additionally generated comb lines and the input laser linewidth dominates the coherence properties.

## 3. Bandwidth scaling simulations

Further bandwidth scaling of frequency combs in Silicon is possible at the systems level by increasing the electro-optical interaction length or by cascading more phase modulators. At the device level, improved RF-coupling (i.e. by use of an appropriate impedance matching network) leads to generation of more phase modulated sidebands and thus more comb lines. This could also be achieved by increasing the doping concentration of the PN junction, the tradeoff being higher free carrier absorption leading to increased absorption losses. Thus, to increase the number of lines, it is preferable to move to a PIN configuration which can support much longer lengths due to reduced absorption losses. The integration of high frequency, high power RF platforms with CMOS based photonics is available for example in MPW [19], however an alternative approach involves the use of 3D integration of the two platforms using Through Silicon Vias (TSV's) or by wire-bonding the two platforms following a system in a package approach.

To investigate the comb generation process, we demonstrate the generation of mostly flat frequency combs by electro-optic simulations on an optimized 10mm long, 450nm wide, 250 nm high Silicon rib waveguide with a shoulder height of 50nm, having abrupt highly doped P++/I/N++ junctions with average doping concentrations of $10^{19}$/cm$^3$ that is routinely used for ohmic contacts with current implantation technology and an intrinsic doping concentration of $10^{15}$/cm$^3$ respectively in a PIN configuration. The device is biased through idealized contacts placed above the highly doped regions that start 50 nm from the rib edge and has a planarizing oxide cladding. These device parameters characterize a narrow-base PIN diode. We model the change in the effective refractive and absorption indices of the intrinsic region due to a sinusoidal modulating input (See Fig. 6) using the Soref-Bennett relations [17] as

$$\Delta n = -\left[ 8.8 \times 10^{-22} \Delta N_e + 8.5 \times 10^{-18} \left( \Delta N_h \right)^{0.8} \right]$$

$$\Delta \alpha = \left[ 8.5 \times 10^{-18} \Delta N_e + 6 \times 10^{-18} \Delta N_h \right]$$

We study the large-signal, electro-optical, AC response of the device to a 4Vpp, 10 GHz sinusoidal input by simulation on a drift-diffusion equation simulator, Silvaco Atlas. Concentration dependent Shockley-Read-Hall recombination, concentration dependent mobility transport, band gap narrowing effects along with Auger recombination and Fermi-Dirac carrier statistics is incorporated in the simulation model.

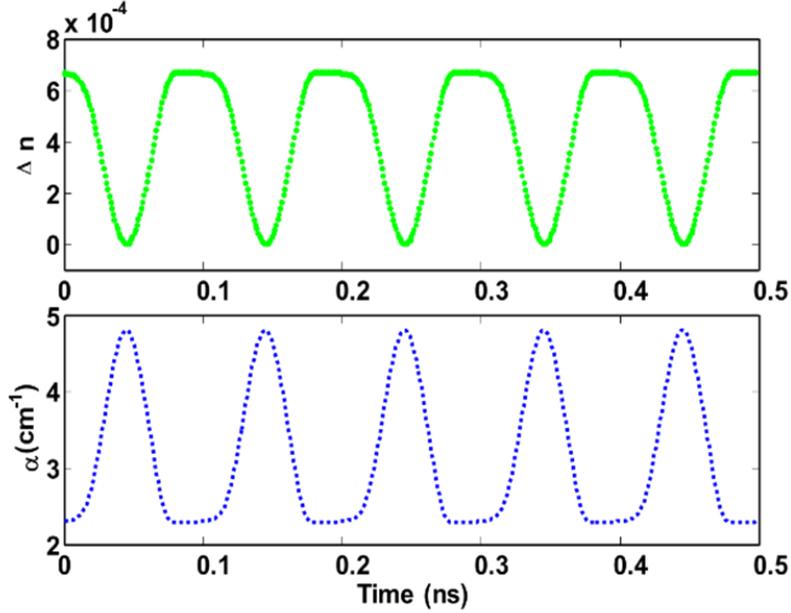

Fig. 6. Simulated change in effective refractive index and absorption generated from an electro-optic drift diffusion simulation for a strongly sinusoidal drive.

The change in refractive index and absorption profiles derived from the simulation (Fig. 6) show mutually out-of-phase saturation regions. The maximum change in refractive index is ~$6.7 \times 10^{-4}$. These saturation regions arise from reduction of charge injection due to the series resistance of the long diode considered in the simulation.

The schematic for mostly-flat frequency comb generation is shown in Fig 7(a). We consider a 10mm long modulator which would correspond to ~ $4V\pi L\pi$, chosen to ensure a high modulation index required for strong phase modulation, whose $L\pi$=2.3mm. This modulator is driven by a 10 GHz microwave source. The loss of our idealized device is found to be ~23 dB.

A ring-resonator based notch filter is used for enhancement of the flatness of the comb. Fig 7(b) shows the spectrum of the generated 10GHz comb. A broad, flat spectrum is seen, however, there is substantial power around the carrier. The reason for that can be seen from the plot of refractive index variation vs time as shown in Fig 6. Since the response to the sinusoid is not strictly sinusoidal, in regions where the refractive index profile is flat, there is no modulation and thus substantially higher fraction of the power rests in and around the carrier. If the response of the modulator was strictly sinusoidal, this effect would not have been observed. The flatness however can be recovered using a simple ring based notch filter by optimally tuning the parameters. The parameters of the notch filter are obtained by solving a simple three-dimensional grid search to determine the coupling coefficient, loss coefficient and ring radius that maximize a quadratic metric of spectral flatness. The additional loss due to the shaping by the ring based notch filter is ~5dB.

Figure 7(b) shows the filter transmission (red) for getting best flatness of the total system. The frequency comb smoothed by the notch filter is observed to have enhanced spectral flatness (Fig. 7(c)).

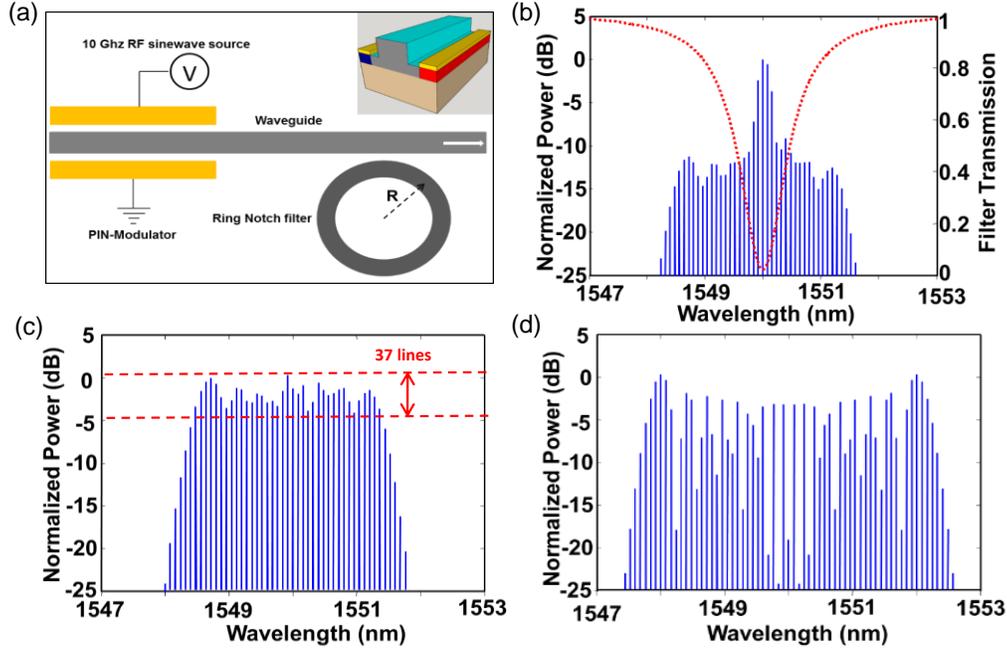

Fig. 7(a) Silicon PIN phase-modulator followed by a ring-resonator based notch filter. (b) Comb lines (blue) from the modulator with notch filter response (red) (c) Flattened Comb lines generated from optimal PIN-modulator device post notch filtering. With 37 usable lines in a 5dB band. (d) Equivalent frequency comb with missing lines generated by pure phase modulation.

Unlike field-effect based modulators that have small and constant time-dependent linear absorption, the time varying free carrier absorption in Silicon charge-injection phase modulators has important consequences for the comb envelope. Fig 7(d) shows the anticipated comb from pure phase modulation with a modulator which has the same level of phase modulation ($\Delta n_{max} = 6.72 \times 10^{-4}$). As is seen, multiple lines are missing, and the flatness is poor. This is expected in the case of pure sinusoidal modulation. However, from our simulations, and from the experimental results (Fig 2) the frequency comb produced by the Silicon phase modulator yields a flat spectrum with no missing lines. The origin of this crucial improvement lies in the amplitude shaping which accompanies phase modulation in these modulators. It has been rigorously shown in theory and experiment [13] [20] that when the amplitude of the time-domain waveform is shaped into a more pulse like structure during sinusoidal phase modulation, the phase modulation itself is equivalent to a time-to-frequency mapping which creates an optical spectrum similar to the time domain waveform. Thus, the shape of the spectrum (Fig 7(c)) is anticipated to be similar to a scaled compliment of the amplitude shaping function in Fig 6. Further details on the effects of amplitude shaping in comb generation can be found in [13] [20].

The cascaded intensity and phase modulator based comb generator systems based on Lithium-Niobate typically have a net loss of around 11dB in conventional operating conditions [14]. Thus, though the experimentally demonstrated system is lossier, it is of comparable performance. However, we now have the advantage of on-chip operation. In addition, we believe that this is not a fundamental issue and it is possible to tune the absorption induced loss by tweaking the doping profile and length of the device appropriately.

### 4. Summary and future work

In this paper, we demonstrated the generation of flat, tunable, high-repetition rate frequency combs in Silicon charge injection based PN-modulators. The modulators were generated in a fully CMOS compatible process. These combs have the unique advantage of continuously tunable center frequency across the C-band and an independently tunable repetition rate (upto 12.5 GHz). We believe that the envelope of the combs can be further enhanced by tuning the doping parameters and introducing asymmetry in both doping concentration and alignment with the optical mode. We also demonstrated through simulations the potential for bandwidth scaling using an optimized Silicon PIN modulator based on similar principles. We demonstrated that the time varying free carrier absorption due to carrier injection, an undesirable effect in data modulators, provides substantial benefits here for improving the spectral flatness and overcoming missing lines present in combs generated using pure phase modulation.

**Funding**

Office of the Principal Scientific Advisor, Govt of India (Prn.SA/ADV/Photonics/2015-16).

**Acknowledgments**

The authors would like to thank Prof SV Raghavan for extensive discussions and initiation of the program. KP Nagarjun would like to thank MHRD for financial support, Sushobhan Avashthi for extensive discussions regarding device behavior and would also like to acknowledge BS Vikram for help with the measurement setup, Abhishek Mishra and Chethan Kumar of DESE, IISc for their help with the simulations, and Nikhil R Kumar for his help with GDS preparation. V R Supradeepa and Shankar Kumar Selvaraja thank MeitY for support through Sir Visvesvaraya young faculty research fellowships.